\documentclass[prd,showpacs,amsmath,amssymb]{revtex4}
\usepackage{graphicx,color,slashbox}
\begin{document}
%\begin{CJK}{GBK}{}

\title{The hidden charm decay of $Y(4140)$ by the rescattering mechanism}

\author{Xiang Liu$^{1,2}$%(ÁõÏè)
}
\email{liuxiang@teor.fis.uc.pt}

\vspace*{1.0cm}

\affiliation{$^1$School of Physical Science and Technology,
Lanzhou University, Lanzhou 730000, China\\
%Department of Physics, Peking University, Beijing
%100871, China\\
$^2$Centro de F\'{i}sica Computacional,
Departamento de F\'{i}sica, Universidade de Coimbra, P-3004-516
Coimbra, Portugal}

\vspace*{1.0cm}

\date{\today}
\begin{abstract}

Assuming that $Y(4140)$ is the second radial excitation of the P-wave
charmonium $\chi_{cJ}^{\prime\prime}$ ($J=0,\,1$), the hidden
charm decay mode of $Y(4140)$ is calculated in terms of the
rescattering mechanism. Our numerical results show that the upper
limit of the branching ratio of the hidden charm decay $Y(4140)\to
J/\psi\phi$ is on the order of $10^{-4}\sim 10^{-3}$ for both of
the charmonium assumptions for $Y(4140)$, which disfavors the
large hidden charm decay pattern indicated by the CDF experiment.
It seems to reveal that the pure second radial excitation of the
P-wave charmonium $\chi_{cJ}^{\prime\prime}$ ($J=0,\,1$) is
problematic.

\end{abstract}

 \maketitle
%\end{CJK}

%%%%%%%%%%%%%%%%%%%%%%%%%%%%%%%%%%
%\section{Introduction}\label{sec1}
%%%%%%%%%%%%%%%%%%%%%%%%%%%%%%%%%

Recently, the CDF experiment announced a new narrow state named
$Y(4140)$ by studying the $J/\psi\phi$ mass spectrum in the exclusive
$B^+\to J/\psi\phi K^+$ process. Its mass and decay width are
$M=4143.0\pm2.9(stat)\pm1.2(syst)$ MeV/c$^2$ and
$\Gamma=11.7^{+8.3}_{-5.0}(stat)\pm 3.7(syst)$ MeV/c$^2$,
respectively \cite{CDF}.

The charmonium-like states discovered in the past six yeasrs include
$X(3872)$, $X(3940)$/$Y(3930)$/$Z(3930)$, $Y(4260)$, $Z(4430)$
etc.. The observation of $Y(4140)$ not only increases the spectrum of
charmonium-like state, but also helps us to further
clarify these observed charmonium-like states.

In our recent work \cite{LZ}, we discussed the various possible
interpretations of the $Y(4140)$ signal. We concluded that
$Y(4140)$ is probably a $D_s^\ast {\bar D}_s^\ast$ molecular state
with $J^{PC}=0^{++}$ or $2^{++}$, while $Y(3930)$ is its $D^\ast
{\bar D}^\ast$ molecular partner, as predicted in our previous work
\cite{lllz}. Later, the author of Ref. \cite{mahajan} also agreed
with the explanation of the $D_s^\ast {\bar D}_s^\ast$ molecular state
for $Y(4140)$ and claimed that hybrid charmonium with $J^{PC} =
1^{-+}$ cannot be excluded. In Ref \cite{nielsen}, they used a
molecular $D_s^\ast {\bar D}_s^\ast$ current with $J^{PC}=0^{++}$
and obtained $m_{D_s^\ast {\bar D}_s^\ast}= (4.14 \pm 0.09)$ MeV,
which can explain $Y(4140)$ as a $D_s^\ast {\bar
D}_s^\ast$ molecular state. The author of Ref. \cite{wang} also
used the QCD sum rules to study $Y(4140)$ and came to a 
different conclusion than that in \cite{nielsen}.

As indicated in our work \cite{LZ}, the study of the decay modes
of $Y(4140)$ is important to test the molecular structure
$D_s^\ast {\bar D}_s^\ast$ of $Y(4140)$. Assuming $Y(3940)$ and
$Y(4140)$ as $D^\ast {\bar D}^\ast$ and $D_s^\ast {\bar D}_s^\ast$
molecular states, respectively, the authors of Ref. \cite{BGL}
calculated the strong decays of $Y(4140)\to J/\psi\phi$ and
$Y(3940)\to J/\psi\omega$ and the radiative decay
$Y(4140)/Y(3940)\to \gamma\gamma$ by the effective Lagrangian
approach. The result of the strong decays of $Y(3940)$ and
$Y(4140)$ strongly supports the molecular interpretation for
$Y(3940)$ and $Y(4140)$.

On the other hand, studying the decay modes with other
structure assignments for $Y(4140)$ will help us to understand
the character of $Y(4140)$ more accurately. Along this line, we
further calculate the hidden charm decay mode of $Y(4140)$
assuming it to be a conventional charmonium state by the rescattering
mechanism \cite{rescattering,mc}.

%It will be helpful to come to a definite conclusion about whether $Y(4140)$
%can be a conventional charmonium state by comparing our result
%with further experimental information.

%%%%%%%%%%%%%%%%%%%%%%%%%%%%%%%%%%%%%%%%%%%%%%%%%%%%%%%%
%\section{The hidden charm decay of $Y(4140)$}\label{sec2}
%%%%%%%%%%%%%%%%%%%%%%%%%%%%%%%%%%%%%%%%%%%%%%%%%%%%%%%%

If $Y(4140)$ is a conventional charmonium state, $Y(4140)$ should
be the second radial excitation of the P-wave charmonium
$\chi_{cJ}^{\prime\prime}$ \cite{LZ}. Its quantum number should be
$J^{P}=0^+,\,1^+,\,2^+$. Since the rather small Q-value for the
decay $B^+\to K^+ Y(4140)$ favors a low angular momentum
$\ell$ between $K^+$ and $Y(4140)$ more, $Y(4140)$ thus favors a low
quantum number $J$ due to $J =\ell$ . In the following, we focus
on the hidden charm decay of $Y(4140)$ with
$\chi_{c0}^{\prime\prime}\,(J^{P}=0^+)$ and
$\chi_{c1}^{\prime\prime}\,(J^{P}=1^+)$ assumptions.

%\subsection{$Y(4140)\to D_{s}^+D_s^-\to J/\psi \phi$ assuming $J^P=0^+$ for $Y(4140)$}

\begin{figure}[htb]
\begin{center}
\scalebox{0.8}{\includegraphics{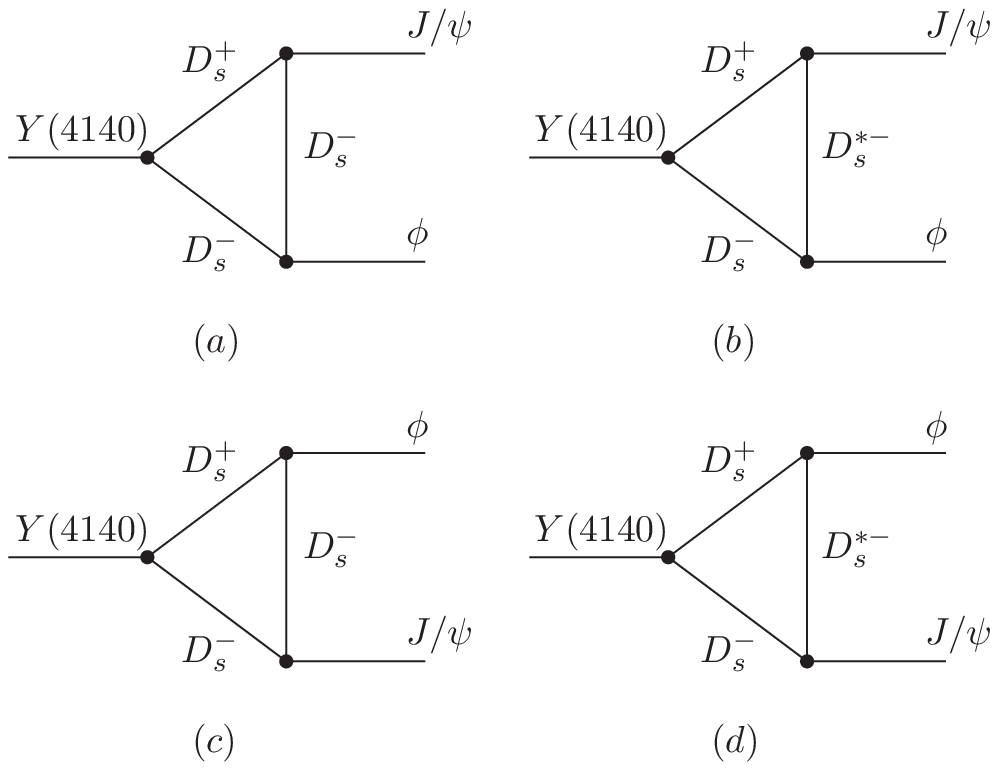}}
\end{center}
\caption{The diagrams for $Y(4140)\to {D}_s^{+}D_s^-\to
J/\psi\phi$ assuming $Y(4140)$ as $\chi_{c0}^{\prime\prime}$
state.}\label{FSI}
\end{figure}

\begin{figure}[htb]
\begin{center}
\scalebox{0.7}{\includegraphics{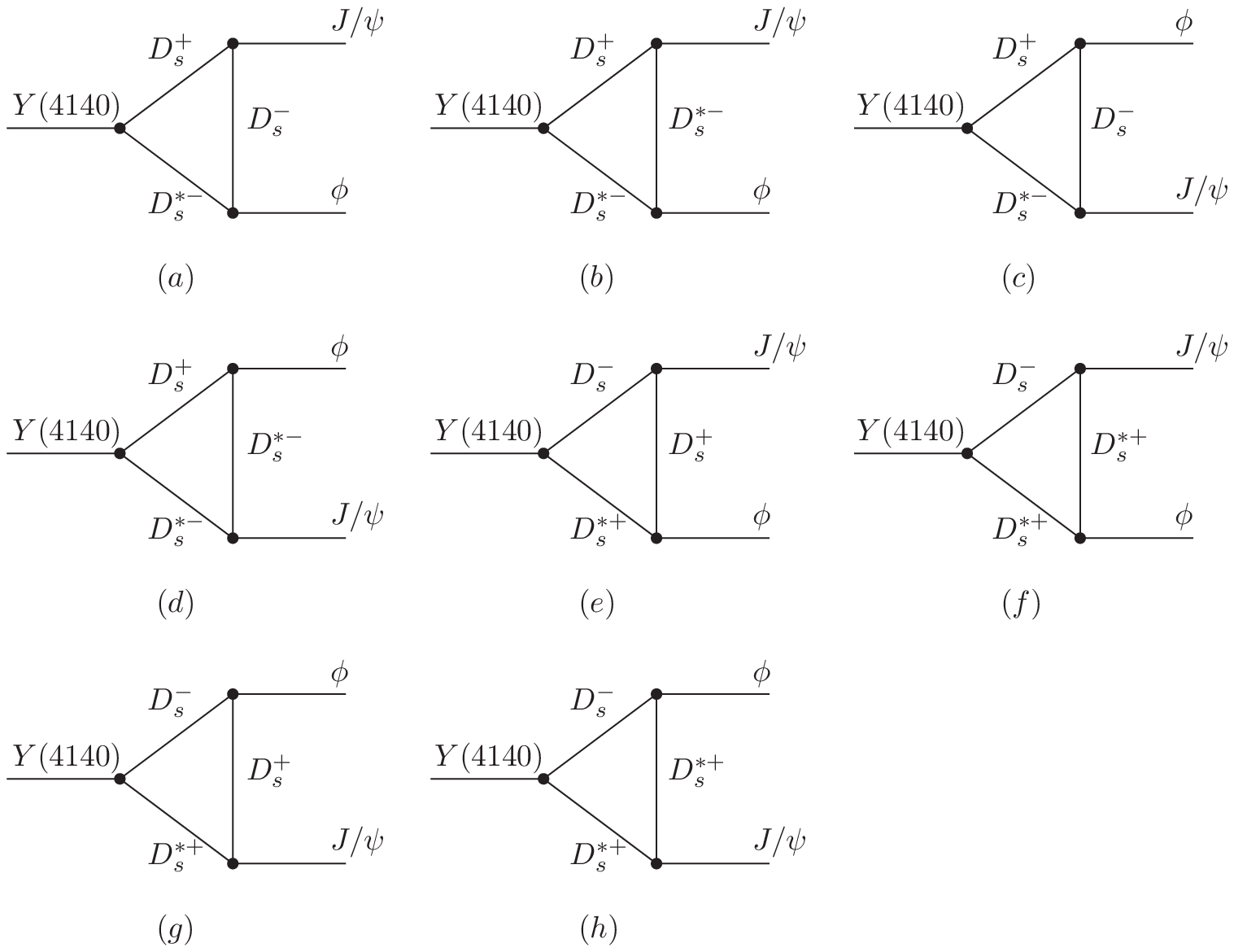}}
\end{center}
\caption{The diagrams for $Y(4140)\to {D}_s^{+}D_s^{*-}+h.c.\to
J/\psi\phi$ assuming $Y(4140)$ as $\chi_{c1}^{\prime\prime}$
state.}\label{FSI-1}
\end{figure}

For the case where $Y(4140)$ is $\chi_{c0}^{\prime\prime}$, the
hidden charm decay $Y(4140)\to J/\psi\phi$ occurs only through
${D}_s^{+}D_s^-$ re-scattering, which is depicted in Fig.
\ref{FSI}. If $Y(4140)$ is $\chi_{c1}^{\prime\prime}$ with
$J^{P}=1^+$, $Y(4140)\to J/\psi\phi$ occurs only via
${D}_s^{+}D_s^{*-}+h.c.$ rescattering, which is shown in Fig.
\ref{FSI-1}.

In Refs. \cite{lagrangian-jpsi,lagrangian-hl,Casalbuoni}, the
effective Lagrangians, which are relevant to the present
calculation, are constructed based on chiral symmetry and 
heavy quark symmetry:
\begin{eqnarray}
\mathcal{L}_{0^{+}DD}&=&g_{Y}\,Y\,D_s^{+}\,{D}_s^{-},\\
\mathcal{L}_{1^{+}DD}&=&g_{Y}Y^{\mu}[D_s^+{{D}_s^{*-}}_{\mu}-D_s^-{D_s^{*+}}_{\mu}
],\\
\mathcal{L}_{_{J/\psi \mathcal{D}\mathcal{D}}}&=&i g_{_{J/\psi
\mathcal{D}\mathcal{D}}}^{} \psi_\mu \left(
\partial^\mu \mathcal{D} {\mathcal{D}}^{\dagger} - \mathcal{D}
\partial^\mu {\mathcal{D}}^{\dagger}
\right),\\
\mathcal{L}_{_{J/\psi \mathcal{D}^* \mathcal{D}}}&=&-g_{_{J/\psi
\mathcal{D}^* \mathcal{D}}}^{} \varepsilon^{\mu\nu\alpha\beta}
\partial_\mu \psi_\nu \left(
\partial_\alpha \mathcal{D}^*_\beta {\mathcal{D}}^{\dagger}
+ \mathcal{D} \partial_\alpha {\mathcal{D}}^{*\dagger}_\beta
\right),\nonumber\\\\
\mathcal{L}_{_{J/\psi \mathcal{D}^* \mathcal{D}^*}}&=&-i
g_{_{J/\psi \mathcal{D}^* \mathcal{D}^*}}^{} \Bigl\{ \psi^\mu
\left(
\partial_\mu \mathcal{D}^{*\nu} {\mathcal{D}}_\nu^{*\dagger} -
\mathcal{D}^{*\nu}
\partial_\mu {\mathcal{D}}_\nu^{*\dagger} \right)
%\nonumber\\&&
+ \left( \partial_\mu \psi_\nu \mathcal{D}^{*\nu} -
\psi_\nu
\partial_\mu \mathcal{D}^{*\nu} \right) {\mathcal{D}}^{*\mu\dagger}  \mbox{} \nonumber\\&& + \mathcal{D}^{*\mu}
\big( \psi^\nu
\partial_\mu {\mathcal{D}}^{*\dagger}_{\nu} - \partial_\mu \psi_\nu {\mathcal{D}}^{*\nu\dagger}
\big) \Bigr\},
\end{eqnarray}
\begin{eqnarray}
\mathcal{L}_{_{\mathcal{D}\mathcal{D}\mathbb{V}}}&=&-ig_{_{\mathcal{D}\mathcal{D}\mathbb{V}}}\mathcal{D}_{i}^{\dagger}{\stackrel{\leftrightarrow}{\partial}}
_{\mu}\mathcal{D}^{j}(\mathbb{V}^{\mu})^{i}_{j},\\
\mathcal{L}_{_{\mathcal{D^{*}}\mathcal{D}\mathbb{V}}}&=&
-2f_{_{\mathcal{D^{*}}\mathcal{D}\mathbb{V}}}\varepsilon_{\mu\nu\alpha\beta}(\partial^{\mu}\mathbb{V}^{\nu})
^{i}_{j}(\mathcal{D}_{i}^{\dagger}
{\stackrel{\leftrightarrow}{\partial}}^{\alpha}\mathcal{D^{*}}^{\beta
j}%\nonumber\\&&
-\mathcal{D^{*}}_{i}^{\beta
\dagger}{\stackrel{\leftrightarrow}{\partial}}^{\alpha}\mathcal{D}^{j}),\\
\mathcal{L}_{_{\mathcal{D^{*}}\mathcal{D^{*}}\mathbb{V}}}
&=&ig_{_{\mathcal{D^{*}}\mathcal{D^{*}}\mathbb{V}}}\mathcal{D^{*}}_{i}^{\nu
\dagger}{\stackrel{\leftrightarrow}{\partial}}_{\mu}\mathcal{D^{*}}_{\nu}^{j}(\mathbb{V}^{\mu})^{i}_{j}%\nonumber\\&&
+
4if_{_{\mathcal{D^{*}}\mathcal{D^{*}}\mathbb{V}}}\mathcal{D^{*}}_{i\mu}^{\dagger}(\partial^{\mu}\mathbb{V}^{\nu}-\partial^{\nu}
\mathbb{V}^{\mu})^{i}_{j}
\mathcal{D^{*}}_{\nu}^{j},\label{lagrangian}
\end{eqnarray}
where $\mathcal{D}$ and $\mathcal{D^*}$ are the pseudoscalar and
vector heavy mesons, respectively, i.e.,
$\mathcal{D^{(*)}}$=(($\bar{D}^{0})^{(*)}$, $(D^{-})^{(*)}$,
$(D_{s}^{-})^{(*)}$). $\mathbb{V}$ denotes the nonet vector meson
matrices.
%\begin{eqnarray}
%\mathbb{V}&=&\left(\begin{array}{ccc}
%\frac{\rho^{0}}{\sqrt{2}}+\frac{\omega}{\sqrt{2}}&\rho^{+}&K^{*+}\\
%\rho^{-}&-\frac{\rho^{0}}{\sqrt{2}}+\frac{\omega}{\sqrt{2}}&
%K^{*0}\\
%K^{*-} &\bar{K}^{*0}&\phi
%\end{array}\right).
%\end{eqnarray}
The values of the coupling constants are \cite{HY-Chen}
$g_{_{DDV}}=g_{_{D^{*}D^{*}V}}=\frac{\beta g_{_{V}}}{\sqrt{2}}$,
$f_{_{D^{*}DV}}=\frac{f_{_{D^{*}D^{*}V}}}{m_{_{D^*}}}=\frac{\lambda
g_{_{V}}}{\sqrt{2}}$, $g_{_{V}}=\frac{m_{_{\rho}}}{f_{\pi}}$,
where $f_{\pi}=132$ MeV, $g_{_{V}},\;\beta$ and $\lambda$ are the
parameters in the effective chiral Lagrangian that describes the
interaction of the heavy mesons with the low-momentum light vector
mesons \cite{Casalbuoni}. Following Ref. \cite{Isola}, we take
$g=0.59$, $\beta=0.9$ and $\lambda=0.56$. Based on the vector
meson dominance model and using the leptonic width of $J/\psi$, the
authors of Ref. \cite{Achasov} determined ${g_{_{J/\psi
\mathcal{D} \mathcal{D}}}^2}/{(4\pi)}=5$. As a consequence of the
spin symmetry in the heavy quark effective field theory,
$g_{_{J/\psi \mathcal{D}{\mathcal{D}}^{*}}}$ and $g_{_{J/\psi
\mathcal{D}^{*}{\mathcal{D}}^{*}}}$ satisfy the relations:
$g_{_{J/\psi \mathcal{D}{\mathcal{D}}^{*}}}={g_{_{J/\psi
\mathcal{D}{\mathcal{D}}}}}/{m_{_{D}}}$ and $g_{_{J/\psi
\mathcal{D}^{*}{\mathcal{D}}^{*}}}=g_{_{J/\psi
\mathcal{D}{\mathcal{D}}}}$ \cite{JPsi-relation}.

Since the contributions from Fig. \ref{FSI} (c) and (d) are the
same as those corresponding to Fig. \ref{FSI-1} (a) and (b),
respectively, the total decay amplitude of
$Y(4140)\to D_s^{+}{D}_s^{-}\to J/\psi\phi$ can be expressed as
\begin{eqnarray}
\mathcal{M}^{(J^{P}=0^+)}&=&2[\mathcal{A}_{1-a}+\mathcal{A}_{1-b}],
\end{eqnarray}
where one formulates the amplitudes of $\mathcal{A}_{1-a}$ and
$\mathcal{A}_{1-b}$ by Cutkosky cutting rule
\begin{eqnarray}
\mathcal{A}_{1-a}&=&\frac{1}{2}\int\frac{d^{3}p_{1}}{(2\pi)^{3}2E_{1}}\frac{d^{3}p_{2}}{(2\pi)^{3}2E_{2}}
%\nonumber\\&&\times
(2\pi)^{4}\delta^{4}(m_{_Y}-p_{1}-p_{2})
[ig_{Y}]\nonumber\\&&\times\Big[-g_{_{J/\psi DD}}i(p_{1}-q)\cdot
\varepsilon_{J/\psi}\Big]%\nonumber\\
%&&\times
\Big[i\;g_{_{DDV}}(q+p_2)\cdot
\epsilon_{\phi}\Big]\nonumber\\&&\times\bigg[\frac{i}{q^2
-m_{D_s}^{2}}\bigg]\mathcal{F}^{2}(m_{D_s},q^2),
\end{eqnarray}

\begin{eqnarray}
\mathcal{A}_{1-b}&=&\frac{1}{2}\int\frac{d^{3}p_{1}}{(2\pi)^{3}2E_{1}}\frac{d^{3}p_{2}}{(2\pi)^{3}2E_{2}}%\nonumber\\&&
%\times
(2\pi)^{4}\delta^{4}(m_{_Y}-p_{1}-p_{2})
[ig_{Y}]\nonumber\\&& \times\Big[i\;g_{_{J/\psi
DD^{*}}}\epsilon_{\mu\nu\kappa\sigma}\varepsilon_{J/\psi}^{\mu}(-i)p_{1}^{\nu}(-i)q^{\sigma}\Big]%\nonumber\\&&\times
\bigg[-2if_{_{D^{*}DV}}\epsilon_{\rho\delta\alpha\beta}ip_{4}^{\rho}\varepsilon_{\phi}^{\delta}
i(p_1^{\alpha}+q^{\alpha})\bigg]\nonumber\\&&
\times\bigg[-g^{\kappa\beta}+\frac{q^{\kappa}q^{\beta}}{m_{D_s^{*}}^{2}}\bigg]\bigg[\frac{i}{q^2
-m_{D_s^{*}}^{2}}\bigg]%\nonumber\\&&\times
\mathcal{F}^{2}(m_{D_s^*},q^2).
\end{eqnarray}

Similarly, we write out the total decay amplitude of $Y(4140)\to
D_s^{+}{D}_s^{*-}+{D}_s^{-}D_s^{*+}\to J/\psi\phi$
\begin{eqnarray}
\mathcal{M}^{(J^{P}=1^+)}&=&2[\mathcal{A}_{2-a}+\mathcal{A}_{2-b}+\mathcal{A}_{2-c}+\mathcal{A}_{2-d}],
\end{eqnarray}
where the pre-factor "2" arises from considering that the
contribution from $D_s^{+}{D}_s^{*-}$ rescattering is the same as
that from ${D}_s^{-}D_s^{*+}$ rescattering. The absorptive
contributions from Fig. \ref{FSI-1} (a)-(d) are, respectively,
\begin{eqnarray}
\mathcal{A}_{2-a}&=&\frac{1}{2}\int\frac{d^{3}p_{1}}{(2\pi)^{3}2E_{1}}\frac{d^{3}p_{2}}{(2\pi)^{3}2E_{2}}
%\nonumber\\&&\times
(2\pi)^{4}\delta^{4}(m_{_Y}-p_{1}-p_{2})
[ig_{Y}\varepsilon_{\xi}]\nonumber\\&&\times\Big[-g_{_{J/\psi
DD}}i(p_{1}-q)\cdot
\varepsilon_{J/\psi}\Big]%\nonumber\\
%&&\times
\Big[-{2}i\;f_{_{D^{*}DV}}\epsilon_{\mu\nu\alpha\beta}ip_{4}^{\mu}\varepsilon_{\phi}^{\nu}(iq^{\alpha}
+ip_{2}^{\alpha})\Big]\nonumber\\&&\times \bigg[-g^{\xi\beta}
+\frac{p_{2}^{\xi}p_{2}^{\beta}}{m_{2}^{2}}\bigg]\bigg[\frac{i}{q^2
-m_{D_s}^{2}}\bigg]\mathcal{F}^{2}(m_{D_s},q^2),\nonumber\\%\nonumber\\
%&=&\int
%d\Omega\frac{|\mathbf{p}_{1}|}{32\pi^{2}m_{_Y}}[2g_{Y}g_{_{J/\psi
%DD}}f_{_{D^{*}DV}}]\nonumber\\&&\times[(2p_{1}-p_{3})\cdot
%\varepsilon_{J/\psi}]
%\epsilon_{\mu\nu\alpha\beta}p_{4}^{\mu}\varepsilon_{\rho}^{\nu}(p_{3}^{\alpha}+p_{2}^{\alpha}-p_{1}^{\alpha})
%\nonumber\\&&\times\bigg[-\varepsilon^{\beta}+p_{2}^{\beta}\frac{p_{2}\cdot
%\varepsilon}{m_{2}^{2}}\bigg]
%\frac{\mathcal{F}^{2}(m_{D},q^2)}{q^{2}-m_{D}^{2}}
\end{eqnarray}

\begin{eqnarray}
\mathcal{A}_{2-b}&=&\frac{1}{2}\int\frac{d^{3}p_{1}}{(2\pi)^{3}2E_{1}}\frac{d^{3}p_{2}}{(2\pi)^{3}2E_{2}}%\nonumber\\&&
%\times
(2\pi)^{4}\delta^{4}(m_{_Y}-p_{1}-p_{2})
[ig_{Y}\varepsilon_{\xi}]\nonumber\\&& \times\Big[i\;g_{_{J/\psi
DD^{*}}}\epsilon_{\mu\nu\kappa\sigma}\varepsilon_{J/\psi}^{\mu}(-i)p_{1}^{\nu}(-i)q^{\sigma}\Big]\nonumber\\&&\times
\bigg\{-{g_{_{D^{*}D^{*}V}}}
i(q+p_{2})\cdot\epsilon_{\phi}g_{\alpha\beta}%\nonumber\\&&
-4f_{_{D^{*}D^{*}V}}\Big[ip_{4\beta}{\epsilon_{\phi}}_{\alpha}-i
{\epsilon_{\phi}}_{\beta}p_{4\alpha}\bigg]\bigg\}\nonumber\\&&
\times\bigg[-g^{\kappa\beta}+\frac{p_{2}^{\kappa}p_{2}^{\beta}}{m_{2}^{2}}\bigg]
\bigg[-g^{\xi\alpha}+\frac{q^{\xi}q^{\alpha}}{m_{D_s^{*}}^{2}}\bigg]%\nonumber\\&&\times
\bigg[\frac{i}{q^2
-m_{D_s^{*}}^{2}}\bigg]\mathcal{F}^{2}(m_{D_s^*},q^2),
\end{eqnarray}
\begin{eqnarray}
\mathcal{A}_{2-c}&=&\frac{1}{2}\int\frac{d^{3}p_{1}}{(2\pi)^{3}2E_{1}}\frac{d^{3}p_{2}}{(2\pi)^{3}2E_{2}}
%\nonumber\\&& \times
(2\pi)^{4}\delta^{4}(m_{_Y}-p_{1}-p_{2})
[ig_{Y}\varepsilon_{\xi}]\nonumber\\&&
\times\bigg[{g_{_{DDV}}}i(q-p_{1})\cdot\varepsilon_{\phi}\bigg]%\nonumber\\&&
%\times
\Big[ig_{_{J/\psi DD^{*}}}
\epsilon_{\mu\nu\alpha\beta}\varepsilon_{J/\psi}^{\mu}iq^{\nu}(-i)p_{2}^{\beta}\Big]\nonumber\\&&
\times\bigg[-g^{\xi\alpha}+\frac{p_{2}^{\xi}p_{2}^{\alpha}}
{m_{D_s^*}^{2}}\bigg]\bigg[\frac{i}{q^2
-m_{D_s}^{2}}\bigg]\mathcal{F}^{2}(m_{D_s},q^2),\nonumber\\
\end{eqnarray}
\begin{eqnarray}
\mathcal{A}_{2-d}&=&\frac{1}{2}\int\frac{d^{3}p_{1}}{(2\pi)^{3}2E_{1}}\frac{d^{3}p_{2}}{(2\pi)^{3}2E_{2}}%\nonumber\\&&
%\times
(2\pi)^{4}\delta^{4}(m_{_Y}-p_{1}-p_{2})
[ig_{Y}\varepsilon_{\xi}]\nonumber\\&&
\times\bigg[-2if_{_{D^{*}DV}}\epsilon_{\mu\nu\alpha\beta}ip_{3}^{\mu}\varepsilon_{\phi}^{\nu}
i(q^{\alpha}-p_{1}^{\alpha})\bigg]%\nonumber\\&&\times
\Big\{-g_{_{J/\psi
D^{*}D^{*}}}\Big[iq^{\kappa}\varepsilon_{J/\psi}^{\sigma}+ip_{2}^{\sigma}\varepsilon_{J/\psi}^{\kappa}\nonumber\\&&
\times+i(p_{2}+q)\cdot\varepsilon_{J/\psi}g^{\kappa\sigma} \Big]
\Big\}\bigg[-g^{\xi}_{\kappa}+\frac{{p_{2}}_{\kappa}p_{2}^{\xi}}{m_{D_s^*}^{2}}\bigg]
%\nonumber\\&&\times
\bigg[-g^{\beta}_{\sigma}+\frac{q_{\sigma}q^{\beta}}{m_{D^{*}}^{2}}\bigg]\bigg[\frac{i}{q^2
-m_{D_s^{*}}^{2}}\bigg]\mathcal{F}^{2}(m_{D_s^*},q^2).\nonumber\\
\end{eqnarray}

In the expressions above for the decay amplitudes,
%$q^{2}=m_{1}^{2}+m_{3}^{2}-2E_{1}E_{3}+2|\mathbf{p}_{1}||\mathbf{p}_{3}|\cos\theta$.
form factors $\mathcal{F}^{2}(m_{i},q^2)$ etc. compensate for the
off-shell effects of the mesons at the vertices and are written as
%\begin{eqnarray}
$\mathcal{F}^{2}(m_{i},q^2)=\bigg(\frac{\Lambda^{2}-m_{i}^2
}{\Lambda^{2}-q^{2}}\bigg)^2$,
%\end{eqnarray}
where $\Lambda$ is a phenomenological parameter. As $q^2\to 0$, the
form factor becomes a number. If $\Lambda\gg m_{i}$, it becomes
unity. As $q^2\rightarrow\infty$, the form factor approaches
zero. As the distance becomes very small, the inner structure
manifests itself, and the whole picture of hadron interaction
is no longer valid. Hence, the form factor vanishes and plays a
role in cutting off the end effect. The expression of $\Lambda$ is
defined as $\Lambda(m_{i})=m_{i}+\alpha \Lambda_{QCD}$
\cite{HY-Chen}. Here, $m_{i}$ denotes the mass of exchanged meson,
$\Lambda_{QCD}=220$ MeV, and $\alpha$ denotes a phenomenological
parameter in the rescattering model.

\iffalse Because the $\rho$ meson is a broad resonance with
$\Gamma_{\rho}\sim 150$ MeV, the decay width of $X(3872)\to
D^{0}\bar{D}^{*0}+\bar{D}^{0}D^{*0}\to J/\psi\rho$ is written as
\begin{eqnarray}
\Gamma=\int^{(M_{_{X(3872)}}-m_{J/\psi})^{2}}_{0}{\rm{d}}s
f(s,m_{\rho},\Gamma_{\rho})\frac{|\mathbf{k}||\mathcal{M}(m_{\rho}\to
\sqrt{s})|^{2}}{24\pi M^{2}_{_{X(3872)}}},\nonumber
\end{eqnarray}
where the Breit-Wigner distribution function
$f(s,m_{\rho},\Gamma_{\rho})$ and the decay momentum
$|\mathbf{k}|$ are
\begin{eqnarray*}
&f(s,m_{\rho},\Gamma_{\rho})=\frac{1}{\pi}
\frac{m_{\rho}\Gamma_{\rho}}{(s-m_{\rho}^{2})^{2}+m_{\rho}^{2}\Gamma_{\rho}^{2}},\\
&|\mathbf{k}|=\frac{\sqrt{[M_{_{X(3872)}}^{2}-(\sqrt{s}+m_{J/\psi})^{2}][M_{_{X(3872)}}^{2}-(\sqrt{s}-m_{J/\psi})^{2}]}}{2M_{_{X(3872)}}}\;.
\end{eqnarray*}\fi

%%%%%%%%%%%%%%%%%%%%%%%%%%%%%%%%%%%%%%%%%%%%%
%\section{Numerical results}\label{sec3}
%%%%%%%%%%%%%%%%%%%%%%%%%%%%%%%%%%%%%%%%%%%%

By fitting the central value of the total width of $Y(4140)$ (11.7
MeV), we obtain the coupling constant $g_{Y}$ in Eq.
(\ref{lagrangian})
\begin{eqnarray*}
g_{Y}=\left\{\begin{array}{cc}
 2.79 \;\mathrm{GeV}, \quad&\mathrm{for}\quad\chi_{c0}^{\prime\prime}\,,\\
 2.65\;\mathrm{GeV},\quad&\mathrm{for}\quad\chi_{c1}^{\prime\prime}\,,
\end{array}\right.
\end{eqnarray*}
where we approximate $D_s^+D_s^-$ and
${D}_s^{+}D_s^{*-}+h.c.$ as the dominant decay mode of $Y(4140)$
when assuming $Y(4140)$ to be $\chi_{c0}^{\prime\prime}$ and
$\chi_{c1}^{\prime\prime}$, respectively. In this way, we can
extract the upper limit of the value of the coupling constant
$g_Y$, which further allows us to obtain the upper limit of the
hidden charm decay pattern of $Y(4140)$.

\begin{figure}[htb]
\begin{center}
\scalebox{0.75}{\includegraphics{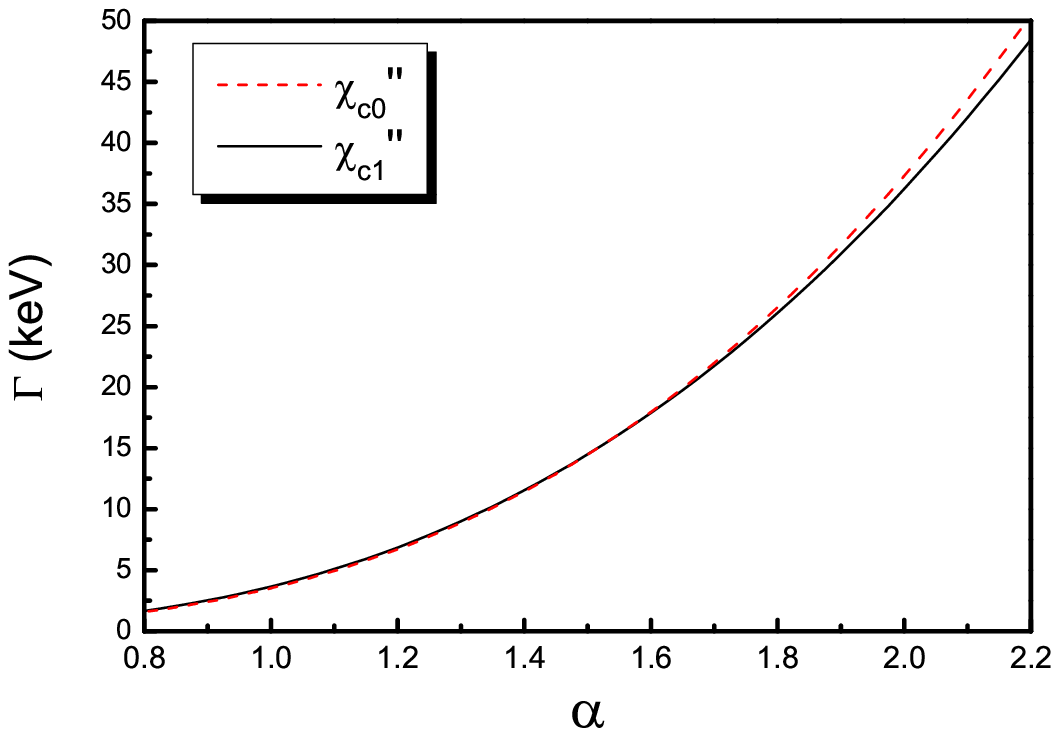}}
\end{center}
\caption{The variation of $\Gamma[Y(4140)\to J/\psi\phi]$ assuming
$Y(4140)$ as $\chi_{c0}^{\prime\prime}$ and
$\chi_{c1}^{\prime\prime}$ states to $\alpha$.}\label{diagram-1}
\end{figure}

The value of $\alpha$ in the form factor is usually of order unity
\cite{HY-Chen}. In this work, we take the range of $\alpha=0.8\sim
2.2$. The dependence of the decay widths of
$Y(4140)(\chi_{c0}^{\prime\prime})\to D_s^{+}{D}_s^{-}\to
J/\psi\phi$ and $Y(4140)(\chi_{c1}^{\prime\prime})\to
D_s^{+}{D}_s^{*-}+h.c.\to J/\psi\phi$ on $\alpha$ is presented in
Fig. \ref{diagram-1}.

In Table \ref{numerical}, we list the typical values of the
branching ratios of $Y(4140)(\chi_{c0}^{\prime\prime})\to
D_s^{+}{D}_s^{-}\to J/\psi\phi$ and
$Y(4140)(\chi_{c1}^{\prime\prime})\to D_s^{+}{D}_s^{*-}+h.c.\to
J/\psi\phi$ when taking different $\alpha$.

\begin{widetext}
\begin{center}
\begin{table}[h]
\begin{tabular}{c|cccccccccc}
\hline \backslashbox{$Y(4140)$}{$\alpha$}& 0.8 & 1.0 & 1.2 & 1.4 &
1.6 & 1.8&2.0&2.2 \\ \hline\hline

$\chi_{c0}^{\prime\prime}$&$1.3\times10^{-4}$&$3.0\times10^{-4}$&$5.7\times10^{-4}$&$9.8\times10^{-4}$&$1.5\times10^{-3}$&$2.3\times10^{-3}$
&$3.2\times10^{-3}$&$4.3\times10^{-3}$\\

$\chi_{c1}^{\prime\prime}$&$1.4\times10^{-4}$&$3.1\times10^{-4}$&$5.9\times10^{-4}$&$9.9\times10^{-4}$&$1.5\times10^{-3}$&$2.2\times10^{-3}$
&$3.1\times10^{-3}$&$4.1\times10^{-3}$\\

 \hline\hline
\end{tabular}\caption{The typical values of the branching ratio of $Y(4140)\to
J/\psi\phi$ for different $\alpha$ assuming $Y(4140)$ to be
$\chi_{c0}^{\prime\prime}$ and $\chi_{c1}^{\prime\prime}$.}
\label{numerical}
\end{table}
\end{center}
\end{widetext}

%%%%%%%%%%%%%%%%%%%%%%%%%%%%%%%%%%%%%%%%%%%%%
%\section{Short summary}\label{sec4}
%%%%%%%%%%%%%%%%%%%%%%%%%%%%%%%%%%%%%%%%%%%%%%%%%%%

In summary, in this paper, we discuss the hidden charm decay of
$Y(4140)$ newly observed by the CDF experiment when assuming
$Y(4140)$ as $\chi_{c0}^{\prime\prime}$ and
$\chi_{c1}^{\prime\prime}$.

According to the rescattering mechanism \cite{rescattering,mc},
the hidden charm decay mode $J/\psi\phi$ occurs via
$D_{s}^+D_{s}^-$ and $D_{s}^{+}D_{s}^{*-}+h.c.$, respectively
corresponding to $\chi_{c0}^{\prime\prime}$ and
$\chi_{c1}^{\prime\prime}$ assumptions for $Y(4140)$. Our
numerical results indicate that the upper limit of the order of
magnitude of the branching ratio of $Y(4140)\to J/\psi\phi$ is
$10^{-4}\sim 10^{-3}$ for both of the assumptions for $Y(4140)$,
which is consistent with the rough estimation indicated in Ref.
\cite{LZ}. Here $Y(4140)$ lies well above the open charm decay
threshold. A charmonium with this mass would decay into an open
charm pair dominantly. The branching fraction of its hidden charm
decay mode $J/\psi \phi$ is expected to be small.

Such small hidden charm decay disfavors the large hidden charm
decay pattern of $Y(4140)$ announced by the CDF experiment
\cite{CDF}, which further supports that explaining $Y(4140)$ as
the pure second radial excitation of the P-wave charmonium
$\chi_{cJ}^{\prime\prime}$ is problematic \cite{LZ}.

We encourage further experimental measurement of the decay
modes of $Y(4140)$, which will enhance our understanding of the
character of $Y(4140)$.

%%%%%%%%%%%%%%%%%%%%%%%%%%%%%%%%
\section*{Acknowledgments}
%%%%%%%%%%%%%%%%%%%%%%%%%%%%%%%%

X.L. would like to thank Prof. Shi-Lin Zhu for useful discussion.
This project was supported by the National Natural Science Foundation
of China under Grant 10705001, the
\emph{Funda\c{c}\~{a}o para a Ci\^{e}ncia e a Tecnologia} \/of the
\emph{Minist\'{e}rio da Ci\^{e}ncia, Tecnologia e Ensino Superior}
\/of Portugal, under contract SFRH/BPD/34819/2007 and in part by A Foundation for the Author of National Excellent Doctoral Dissertation of P.R. China (FANEDD).

\end{document}